\documentclass[prc,showpacs,floatfix]{revtex4}
\usepackage{bm}
\usepackage{graphicx}
\begin{document}
\preprint{MKPH-T-03-1}
\title{
{\bf Low-energy scattering and photoproduction 
of $\eta$-mesons on three-body nuclei}}
\author{A. Fix\footnote{On leave of 
absence from Tomsk Polytechnic University, 
634034 Tomsk, Russia} 
and H. Arenh\"ovel}
\affiliation{
Institut f\"ur Kernphysik,
Johannes Gutenberg-Universit\"at Mainz, D-55099 Mainz, Germany}
\date{\today}
\begin{abstract}
The optical potential approach for low-energy scattering of $\eta$-mesons 
on three-body nuclei is compared to an exact treatment of the $\eta\,3N$ 
system using four-body scattering theory with separable interactions in 
$s$-waves only. The higher-order terms including the interaction of the 
struck nucleon with the surrounding nuclear medium and virtual target 
excitations in between successive $\eta$-scatterings are found to cause 
important corrections. Effects of final state interaction in 
$\eta$-photoproduction on $^3$H and $^3$He are also studied and sizable 
contributions beyond the optical model approach are found.
\end{abstract}

\maketitle


\section{Introduction}

During the last 10 years much effort has been devoted to the study of 
the interaction of an $\eta$-meson 
with very light nuclei. The attention to this area, called
primarily by the pioneering work of \cite{Ued91,Wilk93}, arises from the 
distinctive features of the $\eta$-nuclear system at low energies. 
In more detail:

(i) The $\eta N$ interaction is characterized by the $S_{11}(1535)$ resonance 
located near zero $\eta N$ kinetic energy. 
As a consequence, the $s$-wave part of the $\eta N$
interaction is attractive and rather large near threshold. 
This considerable 
attraction which is assumed to be coherently enhanced in nuclei 
has led to speculations about the existence of $\eta$-nuclear bound states 
which may be formed already in $A =3$ nuclei. 
Although a calculation using an energy 
independent $\eta A$ potential has confirmed this
hypothesis \cite{Wilk93}, 
more sophisticated investigations~\cite{Wyc95,FiAr02} 
have shown that the $\eta N$ interaction is unlikely 
to yield a bound $\eta\,3N$ system even with a relatively large 
real part of the scattering length ${\cal R}e\,a_{\eta N}$\,=\,0.75\,fm. 
The pole of the scattering amplitude
"recedes" to the nonphysical sheet generating an $s$-wave virtual
state. It is important that apparently the pole is located close to the
scattering threshold, resulting in a strong influence on  
low-energy scattering and production processes with $\eta$-mesons. 

(ii) Concerning the formal aspects, the $S_{11}(1535)$ 
resonance, dominating the low-energy 
$\eta N$ interaction, must distort the transparent connection between the 
$\eta N$ and $\eta A$ scattering amplitudes. 
This connection is well established in the pion-nuclear
case within the local-density limit where the equivalent 
optical potential is related in a simple fashion to the elementary $\pi N$
amplitude \cite{ErEr66} (except for real absorption of pions on  
few-nucleon clusters). 
The physical basis of this fact, giving rise to the so-called 
impulse approximation of the optical 
potential, is a large internucleon  
separation distance compared to the range of the $\pi N$ interaction. On the
contrary, due to the resonance pole in the $\eta N$ interaction, the latter 
must be sizably influenced by the nuclear environment. Indeed, 
using $\Gamma=75$ MeV 
for the $S_{11}$(1535) width near the $\eta N$ threshold, we obtain 
for the collision time  
$\Delta t = 2\hbar/\Gamma\approx 1.7\cdot 10^{-23}s$  
which exceeds the time $\Delta t = \hbar/m_\pi \approx 5\cdot 10^{-24}s$ 
associated with the 
pion-exchange $NN$ interaction. Therefore, the validity of the
simplest optical potential for the $\eta A$ interaction 
is expected to be doubtful, and more rigorous models have to be used.

The purpose of the present paper is to explore the 
interaction of $\eta$-mesons with three-body nuclei, a problem which 
can be solved exactly using methods developed within 
four-body scattering theory. 
At the same time, the generalization of the results obtained in this way 
to heavier nuclei seems to be more justifiable than in the deuteron case, 
where the two nucleons are strongly kinematically correlated and 
very weakly bound. 
Our intention is to analyze the quality of the first-order optical potential 
for the $\eta\,3N$ interaction using as a reference the exact four-body 
calculation. Since we have no way of direct fitting $\eta$-nuclear
scattering cross sections  
using a phenomenological potential model, the information on
the low-energy $\eta$-nuclear interaction stems entirely 
from the assumed properties of the $\eta N$ interaction and depends strongly 
on the model linking these two processes. For this reason, 
a thorough microscopic approach to the $\eta$-nuclear dynamics 
becomes particularly important. 
On the other hand, rather complex mathematical infrastructure of the 
four-body scattering theory prevents to some extent a simple interpretation 
of the results. Therefore, we first will clarify the question, whether  
the $\eta$-nuclear 
interaction can be adequately described in terms of the simplest optical 
potential. Furthermore, the comparison of the four-body 
results with those obtained using less rigorous but very tractable 
approaches, such as the lowest-order optical potential, 
may be very fruitful in understanding the $\eta\,3N$ interaction mechanism. 

Several aspects concerning the 
accuracy of the simplest optical model for the scattering 
of $\eta$ mesons on $s$-shell nuclei were already discussed  
in Ref.~\cite{Wyc95}. In particular, it was shown that the behavior  
of the $\eta N$ scattering matrix below the free threshold has a crucial 
influence on the results. Here 
we address other questions related to the optical potential approach  
for $\eta\,3N$ scattering, namely:\\
(i) What is the influence of binding of the participating nucleon 
on the elementary scattering process?\\
(ii) What is the relative importance of target excitations in between two 
successive scatterings on different nucleons?\\
(iii) What is the importance of the short-range behavior of the 
nucleon-nucleon potential?

The second part of the paper is devoted to coherent 
$\eta$-photoproduction on three-body nuclei
\begin{equation}\label{eq5}
\gamma +\,^3\mbox{H}/\,^3\mbox{He}\to \eta+\,^3\mbox{H}/\,^3\mbox{He}\,.
\end{equation}
These reactions are of special importance in $\eta$-nuclear 
physics. Firstly, the main driving mechanism of $\eta$-photoproduction, 
the photoexcitation of the $S_{11}(1535)$-resonance,
is well established. This is in contrast to reactions with 
nucleons as incident particles, where the main mechanism connected
with the short-range part of the $NN$ interaction is presumably 
much more complex and as of yet 
not well understood. Secondly, the energy gap between the
coherent and incoherent thresholds, where the $\eta$-yield is free from the
strong incoherent background is about $\Delta E_\gamma$ = 7 MeV, which
is appreciably larger than the one on the deuteron (about 3 MeV). 
This advantage 
has been partially used in a recent $^3$He$(\gamma,\eta)^3$He experiment 
carried out with the TAPS facility operating at MAMI \cite{Pfeif02}. 
Thirdly, the $\eta$-nuclear 
interaction, which is most important in the $s$-wave, must be 
particularly significant in reactions involving nuclei with nonzero spin. 
As a counter example, the reaction
$^4$He$(\gamma,\eta)^4$He, where the $s$-wave in the final state is totally
suppressed, does not show any strong influence of the final state interaction
(FSI). Finally,
the dynamics of the reactions (\ref{eq5}) may be treated within 
a few-body scattering theory, i.e., formally exactly. 
Though the near-threshold $\eta$-photoproduction on three-body nuclei 
was already considered in Ref.~\cite{Shev02} within the so-called 
finite-rank-approximation (FRA), we reexamine it primarily in order to
show the results of the four-body approach for the $\eta\,3N$ interaction in
the final state. 

The outline of the paper is as follows. First, we briefly 
review in Sect.~\ref{sect1} the four-body
formalism which is relevant for the present consideration. 
For the separable representation of the kernels
we use the energy dependent pole expansion (EDPE) method of 
Ref.~\cite{Sof82}. In Sect.~\ref{sect2}, after a short summary of the 
Kerman-McManus-Thaler (KMT) theory,
we discuss the "standard" optical model for the $\eta\,3N$ elastic scattering 
with particular emphasis on the role of the higher-order corrections like  
nucleon-core interaction and virtual target excitations. The
$\eta$-photoproduction on three-body nuclei is presented in Sect.~\ref{sect3} 
where we illustrate the strong effect of the $\eta\,3N$
interaction in the final state. 
In this section we also compare our predictions with 
those given in~\cite{Shev02}. 
The main results are reviewed in the conclusions. 


\section{The four-body approach to $\eta\,3N$ scattering}\label{sect1}

We begin the formal part with a brief review of 
the four-body scattering formalism applied to $\eta\,3N$ scattering. 
Our basic tool for solving the four-body equations is the
quasiparticle method, reduced to a purely separable representation for
the driving two-body potentials and also for the 
subamplitudes in the (3+1) and (2+2) partitions.
The main features of the method were widely presented in the
literature (see e.g.\ Ref.~\cite{Fons86} and references therein). 
In applying this approach to the $\eta\,3N$ problem, 
the relevant formalism is considered in \cite{FiAr02}. 
Within the quasiparticle method, the whole dynamics
is described in terms of the amplitudes $X_{\alpha 1}$ ($\alpha$=1,2,3)
connecting the three quasi-two-body channels characterized by the following 
partitions
\begin{equation}\label{chan}
\alpha=1:\eta+(3N), \quad \quad       
\alpha=2: N+(\eta NN), \quad \quad  
\alpha=3: (\eta N)+(NN)
\end{equation}
with the initial channel $\alpha$=1. 
To be specific, we consider the triton as target. Because 
we neglect coulomb forces and thus isospin invariance holds,  
the channels with $^3$H and $^3$He are 
identical. Since only the energies up to the three-body threshold 
will be considered, we treat the pion energy relativistically
but use nonrelativistic kinematics for nucleons and the $\eta$-meson. 
Furthermore, due to strong dominance of $s$-waves in $NN$ and $\eta N$
scattering, we assume that in the low-energy region  
only the lowest partial wave ($L=0$) in the $\eta\,3N$ 
system has to be taken into account.

The essence of the calculational scheme is the solution of   
the scattering problem for the two- and three-body 
subsystems specified in the partitions~(\ref{chan}). 
For $\alpha=1$ and 2 we deal with interacting three-body systems.
Using separable representations for the $NN$ and $\eta N$ potentials, 
the corresponding scattering amplitudes can be expressed 
in terms of effective quasi-two-body amplitudes $U_{\alpha;ij}(q,q';{\cal E})$ 
which describe the scattering of a particle on 
a two-body cluster (quasiparticle). 
The corresponding states are specified by 
the indices $i,j$ marking the quasiparticles, e.g., 
$i,j\in\{d,N^*\}$ for $\alpha=2$ where the 
$(NN)$ and $(\eta N)$ systems are denoted as $d$ and 
$N^*$, respectively. The notation $N^*$ is associated with the 
$S_{11}(1535)$ resonance which dominates the low energy $\eta N$ interaction. 
For $\alpha=3$ we have two independent two-particle 
subsystems. The relevant amplitudes are also represented in the   
quasi-two-body form $U_{3;ij}(q,q';{\cal E})$ 
with $i,j\in\{d,N^*\}$~\cite{FiAr02}. 

The reduction of the four-body equations to a numerically manageable 
form is achieved by expanding the amplitudes 
$U_{\alpha;ij}$ into separable series of finite rank $N_\alpha$ 
\begin{eqnarray}
\label{eq35}
U_{1;dd}^{(ss')}(q,q';{\cal E})&=&\sum\limits_{l,m=1}^{N_1}
v_{d;l}^{1(s)}(q;{\cal E})
\Theta_{1;lm}({\cal E})v_{d;m}^{1(s')}(q';{\cal E})\,,\\
\label{eq36}
U_{\alpha;ij}^{(s)}(q,q';{\cal E})&=&\sum\limits_{l,m=1}^{N_\alpha}
v_{i;l}^{\alpha(s)}(q;{\cal E})
\Theta^{(s)}_{\alpha;lm}({\cal E})v_{j;m}^{\alpha(s)}(q';{\cal E})\,,
\quad \alpha=2,3\,. 
\end{eqnarray} 
To condense the formulas to follow, we use here a unified notation 
for the vertex functions or form factors $v^\alpha$ in all three 
channels. They are related to the ones introduced in~\cite{FiAr02} as: 
$v^1_{d;n}=u_n$, $v^2_{i;n}=v_{i;n}$, $v^3_{i;n}=w_{i;n}$, $i\in\{d,N^*\}$.  
For the sake of clarity, 
we note that the amplitude $U_1$ of the $(NN)+N$ scattering 
is a $2\times 2$ matrix according to the spin index $s=0,1$ (we need only
spin-doublet $3N$ states). Here the values $s=0,1$ denote the total spin of 
the ($NN$) $s$-wave cluster with isospin $t=1-s$. 
At the same time, in the partitions $\alpha=2,3$,  
we have two one-dimensional amplitudes $U^{(s)}_\alpha$. 
For instance, in the channel $\alpha=2$ the index $s$ 
numerates two independent $\eta NN$ states
with $J^\pi=0^-$ ($s$=0) and $J^\pi=1^-$ ($s$=1), respectively. 

Considering the identity of the nucleons, the $\eta\,3N$ 
problem is reduced to a 
$3\times 3$ set of integral equations in one scalar variable. 
For the transition amplitudes $X_{\alpha 1}$ connecting the channel 1 to the 
channels $\alpha=2$ and 3 we arrive at a coupled set of equations 
\begin{eqnarray}\label{eq10}
&&X_{\alpha 1;nn'}^{(ss')}(p,p';E)=
Z_{\alpha 1;nn'}^{(ss')}(p,p';E)\nonumber \\
&&\phantom{XXXXX}+\sum_{\beta=2,3}\sum\limits_{l,m}\sum\limits_{\sigma=0,1}
\int\limits_0^{\infty}\widetilde 
Z_{\alpha\beta;nl}^{(s\sigma)}(p,p'';E)\,\Theta^{(\sigma)}_{\beta;lm}
\Big(E-\frac{{p''}^2}{2M_\beta}\Big)\,
X_{\beta 1;mn'}^{(\sigma s')}(p'',p';E)\,\frac{p''^2\,dp}{2\pi^2}\,, 
\quad \alpha=2,3\,,
\end{eqnarray}
where $Z_{\alpha\beta}$ and 
$\widetilde Z_{\alpha\beta}$ 
are the effective potentials realized through particle  
exchange between the quasiparticles in the channels 
$\alpha$ and $\beta$. The arguments of the effective propagators 
$\Theta_\alpha$ are the internal energies of the 
corresponding clusters, given in (\ref{chan}). 
In the case $\alpha=3$ it is equal to the sum of the 
c.m.\ kinetic energies in the $\eta N$ and $NN$ subsystems. 
The reduced masses in the three channels read
\begin{equation}\label{19}
M_1=\frac{3M_Nm_\eta}{3M_N+m_\eta}\,,\quad
M_2=\frac{M_N(2M_N+m_\eta)}{3M_N+m_\eta}\,,\quad
M_3=\frac{2M_N(M_N+m_\eta)}{3M_N+m_\eta}\,.
\end{equation}

\begin{figure}
\includegraphics[scale=.85]{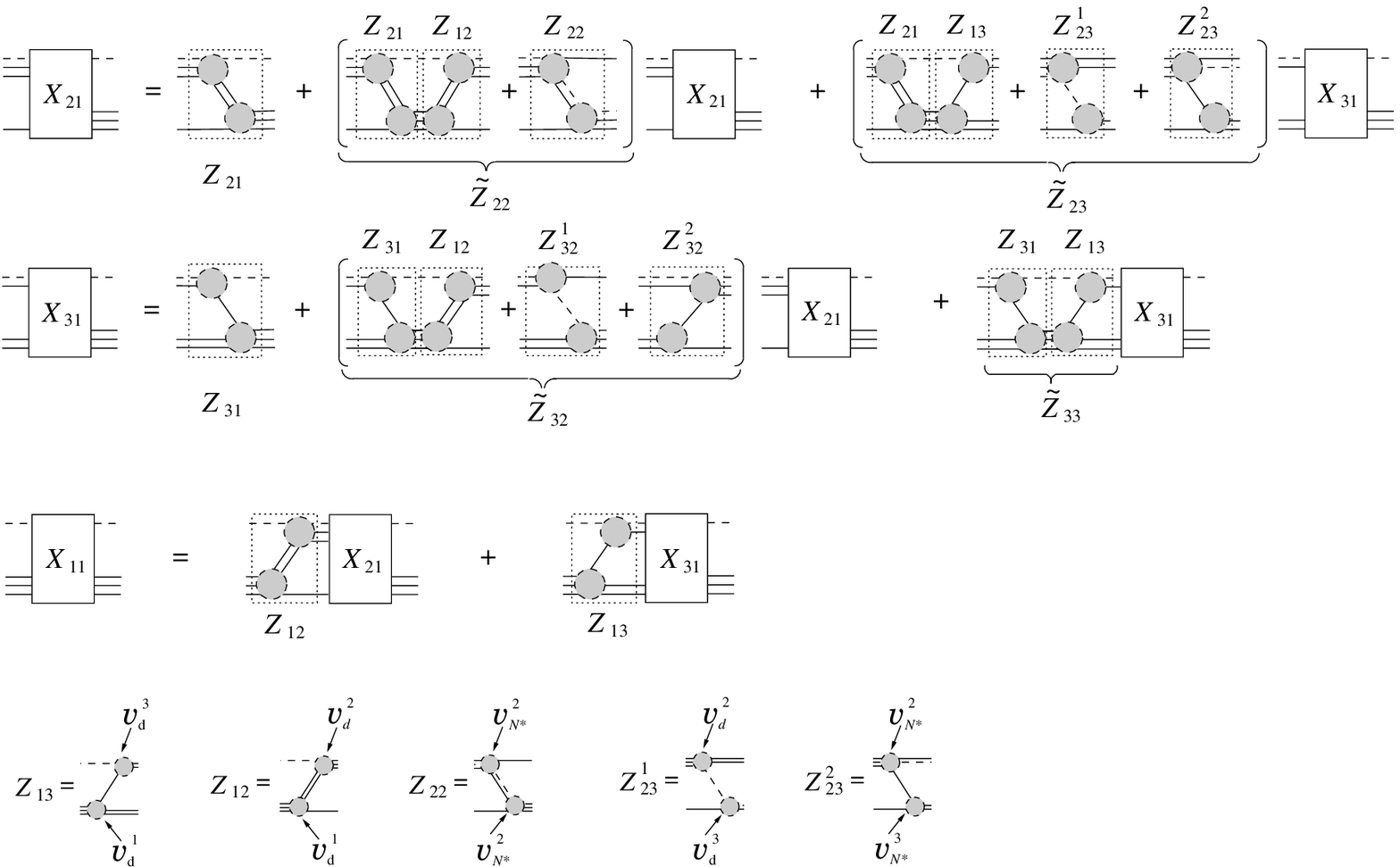}
\caption{
Diagrammatic representation of the coupled integral equations  
(\protect\ref{eq10}) and (\protect\ref{eq20}) for 
the $\eta\,3N$ scattering and the effective potentials $Z_{\alpha\beta}$ 
and $\widetilde Z_{\alpha\beta}$. 
The dashed line represents an $\eta$-meson. The 
lines close together indicate different two- and three-body quasiparticles.  
} 
\label{fig1}
\end{figure}

The equations (\ref{eq10}) are illustrated 
in Fig.~\ref{fig1}, where also 
the structure of the potentials $Z_{\alpha\beta}$ and 
$\widetilde Z_{\alpha\beta}$ is schematically explained. 
The former are expressed in terms of the form factors $v^\alpha_{i;n}$ as 
\begin{equation}\label{Z}
Z^{(ss')}_{\alpha\beta;nn'}(p,p';E)=\frac{\Omega_{ss'}}{2}\sum\limits_{j}
\int\limits_{-1}^{+1} v^{\alpha(s)}_{j;n}(q,E-\frac{p^2}{2M_\alpha})
\,\tau^{(s)}_j\big(E-\frac{p^2}{2M_\alpha}-\frac{q^2}{2\mu_j^\alpha}\big)
\,v^{\beta(s')}_{j;n'}(q',E-\frac{{p'}^2}{2M_\beta})d(\hat{p}\cdot\hat{p}'\,).
\end{equation}
Here, the functions $\tau_j(z)$ are the familiar
quasiparticle propagators appearing in the separable model for  
$NN\,(j=d)$ and $\eta N\,(j=N^*)$ scattering which depend 
on the two-body c.m.\ kinetic energies. 
The corresponding reduced masses $\mu^\alpha_j$ read  
\begin{equation}\label{eq27}
\mu^1_d=\frac{3}{2}M_N\,,\quad
\mu^2_d=\frac{2M_Nm_\eta}{2M_N+m_\eta}\,,\quad
\mu^2_{N^*}=\frac{M_N(M_N+m_\eta)}{2M_N+m_\eta}\,,\quad
\mu^3_d=\frac{M_Nm_\eta}{M_N+m_\eta}\,,\quad
\mu^3_{N^*}=\frac{M_N}{2}\,.
\end{equation}
The spin-isospin coefficients 
are denoted by $\Omega_{ss'}$. Clearly, due to 
the pseudoscalar-isoscalar nature of the $\eta$ meson the spin
$s$ of the $NN$-cluster 
fixes uniquely the order of the spin-isospin coupling of the whole
$\eta\,3N$ configuration. 
The momenta $q$ and $q'$ in (\ref{Z}) are functions  
of the variables $\vec{p},\vec{p}\,'$ and $E$ as given in~\cite{FiAr02}. 
The overall c.m.\ energy $E$ is counted from 
the four-body threshold, i.e.\ $E=W-3M_N-m_\eta$ with $W$ being the 
$\eta\,^3$H invariant mass. Below the first inelastic threshold the obvious 
relation $E\leq -\varepsilon_d$ holds, where $\varepsilon_d$ denotes 
the deuteron binding energy. 
For more details concerning the structure of the potentials 
$\widetilde Z_{\alpha\beta}$ and $Z_{\alpha\beta}$
we refer to Ref.~\cite{FiAr02}. 

Elastic $\eta\,^3$H scattering  
is described by the amplitude $X_{11}$, which is determined by the 
amplitudes $X_{\alpha 1}$ ($\alpha =2,3$) as
\begin{eqnarray}\label{eq20}
X_{11;nn'}^{(ss')}(p,p';E)&=&
\sum_{\alpha=2,3}\sum\limits_{l,m}\sum\limits_{\sigma=0,1}
\int\limits_0^{\infty}
Z_{1\alpha;nl}^{(s\sigma)}(p,p'';E)\,\Theta_{\alpha;lm}^{(\sigma)}
\Big(E-\frac{{p''}^2}{2M_\beta}\Big)\,
X_{\alpha 1;mn'}^{(\sigma s')}(p'',p';E)\,\frac{p''^2dp}{2\pi^2}\,.
\end{eqnarray}

As was already mentioned, the key point of the reduction procedure, leading to
numerically manageable equations (\ref{eq10}), is the 
separable expansion of the subamplitudes (\ref{eq35}) and (\ref{eq36}).
In the present paper we use for this purpose the method of the 
energy dependent pole expansion (EDPE), presented in detail in
\cite{Sof82}. The starting point is the eigenvalue
equation for the vertex functions $v_{i;n}^\alpha(q,{\cal E})$ 
\begin{eqnarray}\label{eq25}
v_{d;n}^{1(s)}(q,B_1)&=&
\frac{1}{\lambda^1_n}\sum\limits_{s'=0,1}\int\limits_0^\infty
V^{(ss')}_{1;dd}(q,q';B_1)\,\tau^{(s')}_d
\big(B_1-\frac{{q'}^2}{2\mu^1_d}\big)
v_{d;n}^{1(s')}(q',B_1)
\frac{q^{\prime 2}dq'}{2\pi^2}\,, \\
\label{eq26}
v_{i;n}^{\alpha(s)}(q,B_\alpha)&=&
\frac{1}{\lambda^{\alpha(s)}_n}
\sum\limits_{j=d,N^*}\int\limits_0^\infty
V^{(s)}_{\alpha;ij}(q,q';B_\alpha)\,\tau_j^{(s)}
\big(B_\alpha-\frac{{q'}^2}{2\mu^\alpha_j}\big)
v_{j;n}^{\alpha(s)}(q',B_\alpha)
\frac{q^{\prime 2}dq'}{2\pi^2}\,, \quad \alpha=2,3\,.
\end{eqnarray}
The explicit expressions for the effective potentials
$V_{\alpha;ij}(q,q';{\cal E})$ are given in Ref.~\cite{FiAr02}. 
The equations (\ref{eq25}) are solved for an arbitrarily fixed energy  
${\cal E}=B_\alpha$. In the actual calculation we have taken 
$B_1=-\varepsilon_{^3\!H}$\,(the triton binding energy) and 
$B_\alpha=-\varepsilon_d$ in the other two channels $\alpha=2,3$.

The extrapolation of the vertices 
$v_{i;n}^\alpha$ onto the whole energy axes is carried out according 
to the expressions 
\begin{eqnarray}\label{eq30}
v_{d;n}^{1(s)}(q,{\cal E})&=&\sum_{s'=0,1}\int\limits_0^\infty
V^{(ss')}_{1;dd}(q,q';{\cal E})
\,\tau_d^{(s')}\big(B_1-\frac{{q'}^2}{2\mu^1_d}\big)
\,v_{d;n}^{(s')}(q',B_1) 
\frac{q^{\prime 2}dq'}{2\pi^2}\,,\\
\label{eq31}
v_{i;n}^{\alpha(s)}(q,{\cal E})&=&\sum_{j=d,N^*}\int\limits_0^\infty
V^{(s)}_{\alpha;ij}(q,q';{\cal E})
\,\tau^{(s)}_j\big(B_\alpha-\frac{{q'}^2}{2\mu^\alpha_j}\big)
\,v^{\alpha(s)}_{j;n}(q',B_\alpha) 
\frac{q^{\prime 2}dq'}{2\pi^2}\,,\quad \alpha=2,3\,.
\end{eqnarray}
The effective EDPE propagators $\Theta_\alpha$ in (\ref{eq35}) 
and (\ref{eq36}) are defined by 
\begin{eqnarray}\label{eq45}
\big(\Theta^{-1}_1({\cal E})\big)_{mn}
&=&\sum_{s=0,1}\int\limits_0^\infty
\big[
v_{d;m}^{1(s)}(q,B_1)\tau^{(s)}_d
\big(B_1-\frac{q^2}{2\mu^1_d}\big)
-v^{1(s)}_{d;m}(q,{\cal E})\tau^{(s)}_d
\big({\cal E}-\frac{q^2}{2\mu^1_d}\big)
\big]
v_{d;n}^{1(s)}(q,{\cal E})\frac{q^2dq}{2\pi^2}\,,\\
\big(\Theta^{(s)-1}_\alpha({\cal E})\big)_{mn}
&=&\sum_{j=d,N^*}\int\limits_0^\infty
\big[
v_{j;m}^{\alpha(s)}(q,B_\alpha)\tau^{(s)}_j
\big(B_\alpha-\frac{q^2}{2\mu^\alpha_j}\big)
-v_{j;m}^{\alpha(s)}(q,{\cal E})\tau^{(s)}_j
\big({\cal E}-\frac{q^2}{2\mu^\alpha_j}\big)
\big]
v_{j;n}^{\alpha(s)}(q,{\cal E})\frac{q^2dq}{2\pi^2}\,,\quad \alpha=2,3\,.
\end{eqnarray}
In the calculation, we use a $6\times 6$ separable representation 
(\ref{eq35}) and (\ref{eq36}) in each partition (\ref{chan}) 
which yields accurate
solutions up to the first inelastic threshold. 

Due to the strong coupling between the
$\eta N$ and $\pi N$ channels in the $S_{11}(1535)$ region, the transitions  
$\eta N \leftrightarrow \pi N$ must in general be taken into account. 
Clearly, the most straightforward way to introduce 
the pion degrees of freedom  
would be to generalize the $\eta\,3N$ four-body equations to include 
the coupled
channels $(\pi\,3N)\leftrightarrow(\eta\,3N)$. But in practice, the
four-body treatment of the $\pi\,3N$ states turns out to be 
very complicated. The reason for this is the appearance of 
moving singularities 
arising near the physical region for the $\pi NN$ amplitudes 
above the three-body threshold. As a result, the separable representation 
of the four-body kernels converges very poorly~\cite{footn1}. 
Therefore, we neglect the channel $\pi\,3N$ 
keeping only the intermediate $\pi N$-"bubbles" in the $S_{11}(1535)$
propagator. The validity of this neglect seems to be doubtful, 
since the $\pi N$ interaction in the second resonance
region is visibly stronger than the $\eta N$ one. 
The crucial point, however, is that 
the two-step process $\eta N\to\pi N\to\eta N$, favoring large momenta 
of the intermediate pion $k_\pi\approx 400$ MeV/c,   
needs two nucleons to be within the range $R=\hbar/k_\pi\approx$ 0.5\,fm. 
Adopting a simple geometric interpretation,  
the corresponding mechanism is associated with a small probability 
\begin{equation}
P=\frac{4}{3}\pi R^3\,\rho_{^3\!H}(0)\approx \frac{1}{10}\,,
\end{equation}
where $\rho_{^3\!H}(r)$ is the $^3$H nucleon density,  
and thus is not expected to be effective for low-energy $\eta\,^3$H 
scattering. 

For the target wave function we take only the $s$-wave part 
\begin{equation}\label{eq70}
\Psi_{M_JM_T}
(\vec{q},\vec{k}\,)=\frac{1}{\sqrt{3}}\left(1-P_{12}-P_{13}\right)
\sum\limits_{s=0,1}\psi^{(s)}(q_1,k_{23})\Big[\big[\frac{1}{2}\times
\frac{1}{2}\big]^{st}\times\frac{1}{2}\Big]
^{\frac{1}{2}M_J\,\frac{1}{2}M_T}\,,
\end{equation}
where the isospin $t=1-s$ and $M_J$ and $M_T$ denote total spin and 
isospin projections, respectively.
The spatial functions $\psi^{(s)}(q_1,k_{23})$ are taken symmetric with 
respect to the nucleons 2 and 3. They are extracted from the bound state 
pole of the $3N$ scattering amplitude, calculated within the three-body model.
The corresponding expression in terms of the $(3N)\to N+(NN)$ vertices 
$v^{1(s)}_{d;1}$ reads 
\begin{equation}\label{eq74}
\psi^{(s)}(q,k)=-Ng_d^{(s)}
\tau^{(s)}_d\big(-\varepsilon_{^3\!H}-\frac{3q^2}{4M_N}\big)
\frac{v^{1(s)}_{d;1}(q,-\varepsilon_{^3\!H})}
{\displaystyle\varepsilon_{^3\!H}+\frac{3q^2}{4M_N}+\frac{k^2}{M_N}}.
\end{equation}
The normalization factor is obtained from the residue of the scattering 
matrix 
\begin{equation}\label{eq76}
N^{-2}=\left.\frac{d\lambda^1_1}{d{\cal E}}\right|_
{{\cal E}=-\varepsilon_{^3\!H}}\,,
\end{equation}
where $\lambda_1^1$ is the first eigenvalue of the kernel $V_{1;dd}\tau_d$
(see Eq.~(\ref{eq25})). Finally, for the $\eta\,^3$H scattering amplitude 
we get 
\begin{equation}\label{eq78}
F_{\eta^3\!H}(k_\eta)=-\frac{\mu_{\eta^3\!H}}{2\pi}\sum\limits_{s,s'=0,1}
X_{11;11}^{(ss')}(k_\eta,k_\eta,;E)\,,
\end{equation}
with the $\eta\,^3$H reduced mass $\mu_{\eta^3\!H}$ 
and the on-shell momentum 
$k_\eta=\big[2\mu_{\eta^3\!H}(E+\varepsilon_{^3\!H})\big]^{1/2}$.

As was already noted, we consider  
the $NN$ and $\eta N$ interactions only in $s$ states. 
For the $NN$ $^1S_0$ and $^3S_1$ configurations 
we adopt a rank-one separable parametrization 
\begin{equation}\label{eq85}
v_{NN}^{(s)}(k,k')=-g_d^{(s)}(k)g_d^{(s)}(k')\,,\quad \mbox{with}\ 
g_d^{(s)}(k)=\sqrt{2}\pi
\sum_{i=1}^6\frac{C_i^{(s)}}{k^2+\beta_i^{(s)2}}\,, \
\quad \mbox{for}\ s=0,1\,,
\end{equation}
where the parameters $C_i^{(s)}$ and $\beta_i^{(s)}$ are 
listed in Ref.~\cite{Zan83}. The index $s=0,1$ refers to the singlet and
triplet states, respectively.
The separable potential (\ref{eq85}) is obtained by fitting the off-shell 
behavior of the Paris $NN$ potential at zero energy and is therefore 
appropriate for processes without target break-up.
The corresponding three-body calculation gives for the triton binding 
energy a reasonable value $\varepsilon_{^3\!H}$ = 8.64 MeV
and describes rather well the $^3$H charge form factor up to
$Q^2$ = 8\,fm$^{-2}$.  

\begin{figure}
\includegraphics[scale=.7]{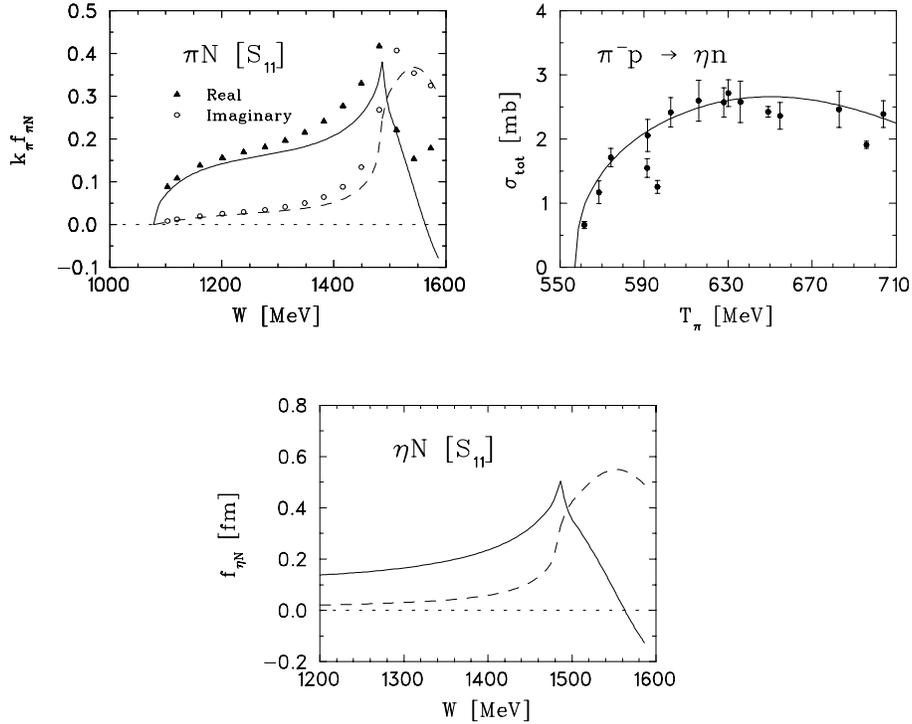}
\caption{
Upper left panel: 
the $S_{11}$ partial wave of the $\pi N$ scattering amplitude
predicted by the parametrization of $S_{11}(1535)$ resonance 
used in the present paper
(see Eq.~(\protect\ref{eq90}) and (\protect\ref{eq102})). Notations:  
solid curve: real part, dashed: imaginary part. Circles and triangles    
represent the result of the VPI-analysis~\protect\cite{Arn85}.
Upper right panel: total $\pi^-p\to\eta n$ cross section. 
The data are taken from the compilation presented in~\protect\cite{BSSN95}.
Lower panel: 
the $\eta N$ off-shell scattering amplitude 
$f_{\eta N}(\vec{q},\vec{q}\,';W)$ at 
$q=q'=0$. Notations as in the upper left panel. 
}
\label{fig2}
\end{figure}

As for the $\eta N$ interaction, we use here the simplest
separable parametrization with the energy-dependent potential
\begin{equation}\label{eq90}
v_{\eta N}(k,k';W)=\frac{g^{(\eta)}_{N^*}(k)g^{(\eta)}_{N^*}(k')}{W-M_0}\,,
\ \quad  \mbox{with}\ \ 
g^{(\eta)}_{N^*}(k)=\frac{g^{(\eta)}_{N^*}}{\sqrt{2\omega_\eta}}
\frac{\beta^{(\eta)2}_{N^*}}{k^2+\beta^{(\eta)2}_{N^*}}\,,
\end{equation}
which gives the familiar isobar ansatz for the meson-nucleon 
amplitude with the bare resonance mass $M_0$. 
In the present paper, the excitation of the $S_{11}(1535)$ resonance is
assumed to be the only mechanism for the meson-nucleon interaction. 
Rather than to investigate the dependence of the results on the $\eta N$
scattering length $a_{\eta N}$ we preferred to choose the parameters in  
(\ref{eq90}) such that the $\eta N$ scattering length  
\begin{equation}\label{eq95}
a_{\eta N}=(0.50+i0.32)\, \mbox{fm}
\end{equation}
is reproduced. 
This value lies approximately "halfway" in the listing 
of various $\eta N$ scattering lengths which can be found 
in the literature (see e.g.\,~\cite{Hai02}). 
It must be noted that the low-energy $\eta$-nuclear interaction depends 
strongly on the continuation of the $\eta N$ 
amplitude to negative kinetic energies 
and hence must be sensitive to the 
amplitudes in the channels coupled to the $\eta N$ one. 
Therefore, we use here the unitary model of Ref.~\cite{BeTa90} where 
three coupled channels $\eta N$, $\pi N$ and $\pi\pi N$ are considered. 
In order to reproduce the value 
(\ref{eq95}) we have slightly changed the set of parameters 
presented in
\cite{BeTa90} in such a manner that the $\pi N\to\pi N$ 
and $\pi N\to\eta N$ scattering data 
are reasonably well described in the region below and just above the $\eta N$
threshold. The results shown in Fig.~\ref{fig2} are obtained with the  
parameter values  
\begin{eqnarray}\label{eq102}
&&g^{(\pi)}_{N^*}=8.898/\sqrt{12\pi}\,, \quad 
\beta^{(\pi)}_{N^*}=404\ \mbox{MeV}\,,\nonumber \\
&&g^{(\eta)}_{N^*}=7.090/\sqrt{4\pi}\,, \quad 
\beta^{(\eta)}_{N^*}=695\ \mbox{MeV}\,, \\
&&M_0=1599\ \mbox{MeV}\,.\nonumber 
\end{eqnarray}
The additional factors in (\ref{eq102}) appear due to different 
normalizations of the meson-nucleon potentials $v_{\pi N}$ and $v_{\eta N}$
used in this work and in \cite{BeTa90}. The parameters of the 
two-pion channel $\pi\pi N$ were taken unchanged from 
Ref.~\cite{BeTa90}.


\section{The optical model for $\eta\,^3 $H scattering}\label{sect2}

We begin the analysis of the optical potential approach by reviewing the
corresponding formalism.  
According to the Watson multiple scattering theory~\cite{GoWa64},
the $\eta$-nuclear
interaction may be treated as a series of $\eta N$ collisions. 
In the present discussion we use the version put 
forward by Kerman-McManus-Thaler (KMT) \cite{KMT59}. 
The corresponding expansion of the scattering operator reads
\begin{equation}\label{eq100}
T=\sum\limits_{i}^A \tau(i)+\sum\limits_{i\ne j}^A \tau(i)G\,\tau(j)+ \cdots
\end{equation}
Here the Green's function $G$ describes the propagation of a free $\eta$ and
$A$ interacting nucleons 
\begin{equation}\label{eq105}
G=\frac{\hat{A}}{E-H_0-V_A}\,,
\end{equation}
where the nuclear potential $V_A$
describes the interactions of the nucleons, 
and the free Hamiltonian $H_0$ includes only the 
kinetic energy operator of meson and nucleons. 
Furthermore, we have included into the Green's function the 
projection operator $\hat{A}$ 
onto the completely antisymmetric nuclear states. 
The scattering matrix $\tau$ in (\ref{eq100}) 
describes the off-shell scattering 
of an $\eta$-meson on a single bound nucleon and obeys the equation 
\begin{equation}\label{eq120}
\tau=v_{\eta N}+v_{\eta N}G\,\tau\,.
\end{equation}
Within the KMT theory the nucleons are treated to be identical 
from the beginning. Therefore 
we have dropped the nucleon index $i$ in (\ref{eq120}).
The formal reduction of the $(A+1)$-body problem 
(\ref{eq100})-(\ref{eq120}) to a two-body scattering problem 
leads for the $T$-matrix to the equation 
\begin{equation}\label{eq125}
T=U+\frac{A-1}{A}\,UGP_0T\,,
\end{equation}
with $P_0$ being the projector onto the nuclear ground state.
The operator $U$ of the equivalent optical potential obeys the equation 
\begin{equation}\label{eq127}
U=U^{(1)}+\frac{A-1}{A}\,U^{(1)}G(1-P_0)U,
\end{equation}
with the driving term $U^{(1)}=A\tau$.

The simplest realization of the multiple scattering formalism 
is mainly based on the following approximations:

(i) Coherent approximation: here one keeps only the
leading term in (\ref{eq127}). The resulting optical potential 
is given by the the ground state expectation value of the 
matrix $\tau$ times the number of nucleons in the nucleus 
\begin{equation}\label{eq135}
U_C(\vec{p},\vec{p}\,';E)=
\langle 0;\vec{p}\,|U^{(1)}(E)|0;\vec{p}\,'\rangle \
=A\langle 0;\vec{p}\,|\tau|0;\vec{p}\,'\rangle \,.
\end{equation}
As may be seen, the restriction to the first-order term $U^{(1)}$ in 
the expansion (\ref{eq127}) neglects the virtual
target excitations in between successive scatterings on different nucleons.  
Of course, as follows from (\ref{eq120}), 
excitations are allowed for when the $\eta$ interacts successively 
with the same 
nucleon. One of the points in favor of the coherent approximation is the 
assumed dominance of the nearest singularity (bound state pole). 
Probably no less important is the orthogonality of the 
nuclear ground state wave function to the excited states, 
which results in a 
reduction of the matrix element $\langle 0|U^{(1)}|n\rangle$ at least at small 
momentum transfers. However, in spite of these reasonable arguments the 
study of $nd$~\cite{AAY64} and $\eta d$~\cite{Wyc01} scattering 
has shown that keeping only the target ground state weakens sizably  
the overall interaction in the system and is therefore a rather poor 
approximation. 

(ii) Impulse approximation: it is considered as a further 
simplification of the approximation (i) and consists in the substitution of the
operator $\tau$ in (\ref{eq135}) by the 
free-space $\eta N$ scattering matrix $t_{\eta N}$ satisfying the equation  
\begin{equation}\label{eq162}
t_{\eta N}=v_{\eta N}+v_{\eta N}G_0t_{\eta N}\,,\quad \mbox{with}\ \ 
G_0=\frac{1}{E-H_0}\,.
\end{equation} 
The resulting optical potential for the $\eta\,^3 $H scattering is then  
\begin{equation}\label{eq165}
U_I(\vec{p},\vec{p}\,';E)=A\langle 0;\vec{p}\,|t_{\eta N}
|0;\vec{p}\,'\rangle 
=A\int\Psi_{^3\!H}^*\big(\vec{q},\vec{k}\,\big)t_{\eta
N}\big(w_{\eta N}(\vec{q}\,)\big)
\Psi_{^3\!H}\big(\vec{q}+\frac{2}{3}(\vec{p}-\vec{p}\,'),\vec{k}\,\big)
\frac{d^3q}{(2\pi)^3}\frac{d^3k}{(2\pi)^3}\,,  
\end{equation}
where the argument $\vec{q}$
of the ground state wave function $\Psi_{^3\!H}$ 
is the relative momentum of the participating nucleon with respect to the
other two nucleons. Thus within this approximation, 
the struck nucleon is bound only before and after the interaction 
with the incident meson but is free during the scattering. The role of the 
surrounding nucleons is to provide only the momentum distribution 
for the active scatterer. 
The impulse approximation is more or less successful for low-energy
pion-nucleus scattering far away from the resonance region~\cite{ErEr66}, 
that is, for light projectiles which interact weakly with the target 
constituents. But its validity may be marginal in the 
$\eta$-nuclear case. The main reason for this fact 
is that the impulse approximation 
breaks down if the projectile is in resonance with the nucleon~\cite{GoWa64}.
As was already noted in the introduction, since the $\eta N$-scattering is
associated with a nonvanishing time delay  
due to the $S_{11}(1535)$ resonance, the interaction of the struck nucleon 
with the remaining ones must generally be important. 

In order to study the quality of the approximations (i) and (ii) for 
the $\eta$-nuclear interaction we have calculated the $\eta \,^3$H 
elastic scattering using the optical 
potentials $U_C$\,(\ref{eq135}) and $U_I$\,(\ref{eq165}).
In each case only the $s$-wave part of the scattering amplitude was
taken into account. The results are obtained by solving 
equation (\ref{eq125}) in momentum space 
\begin{equation}\label{eq175}
T(p,p;E)=U(p,p;E)+\frac{2}{3}\,\frac{\mu_{\eta\,^3\!H}}
{\pi^2}\int\limits_0^\infty
\frac{U(p,p';E)\,T(p',p;E)}{p^2-{p'}^2+i\varepsilon}\ {p'}^2 dp'\,.
\end{equation}  
The numerical difficulties caused by the singularity in the integrand at 
$p'=p$ were eliminated with the help of the Noyes-Kowalski 
trick~\cite{NoKo65}. 

It is worthwhile to note that since the in-medium $\eta N$ 
scattering matrix $\tau$ 
is an $(A+1)$-body operator, its 
full treatment is in general possible only with certain approximations. 
However, in the case of $A=3$, the equation (\ref{eq120}) can be solved 
exactly using the four-body formalism.  
Indeed, this equation represents the reduced
problem where the $\eta$-meson is scattered off only one of 
the nucleons which in turn interacts with the other two nucleons.
Therefore, using the separable representation for the two- and three-body
scattering matrices as described in Sect.~\ref{sect1}, 
the equation (\ref{eq120}) 
can be transformed into the form presented in (\ref{eq10}) and (\ref{eq20}) 
where now we must switch off the $\eta$ exchange 
between the nucleons. In the computation, we simply set the 
potential $V_{2;N^*N^*}$ in Eqs.\,(\ref{eq26}) and (\ref{eq31})
as well as the term $Z^1_{23}(Z^1_{32})$ in the potential 
$\widetilde{Z}_{23}(\widetilde{Z}_{32})$ (see Fig.~\ref{fig1}) 
equal to zero. The matrix $X_{11}$, 
obtained in this way, yields the $s$-wave potential $U_C$ (\ref{eq135}) 
for the $\eta\,^3$H scattering in the form 
\begin{equation}\label{eq160}
U_C(p,p';E)=\sum\limits_{ss'=0,1}X^{(ss')}_{11;11}(p,p';E)\,.
\end{equation}

\begin{figure}
\includegraphics[scale=.7]{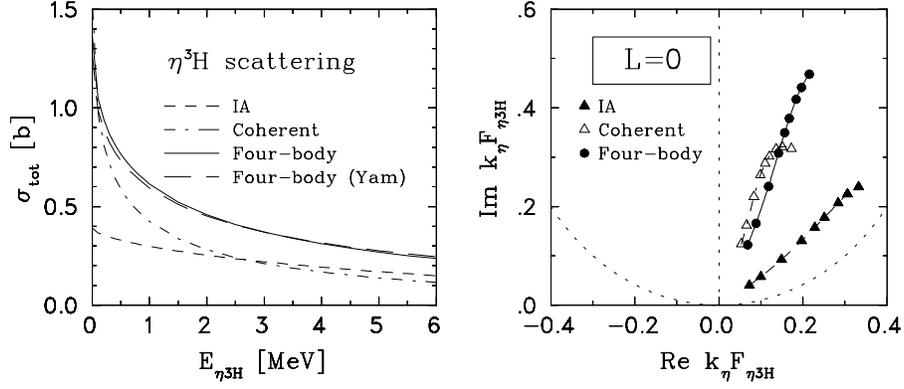}
\caption{Elastic cross section for $\eta\,^3$H scattering 
(left panel) and Argand plot (right panel) of the scattering amplitude. 
The dashed curves (filled triangles on the right panel) 
represent the impulse approximation to 
the first-order optical 
potential (Eq.(\protect\ref{eq165})). In the dash-dotted 
curves (open triangles) 
the medium corrections to the single scattering are taken into account 
(Eq.(\protect\ref{eq135})). 
The solid curves (filled circles)
represent the result of the full four-body calculation. The 
long-dashed curve in the left panel 
is obtained with the Yamaguchi $NN$-potential embedded into the
nuclear sector. In the right panel the circles and triangles indicate 
the following c.m.\ kinetic energies : 
$E_{\eta\,^3\!H}$ = 0.1, 0.2, 0.5, 1.0, 1.5, 2.0, 3.0, 4.0, 
and 6.0 MeV.
} \label{fig3}
\end{figure}

In Fig.~\ref{fig3} we compare the results of the approximations (\ref{eq135})
 and (\ref{eq165})
with those given by the four-body theory where the 
$\eta\,3N$ multiple scattering series (\ref{eq100}) is summed exactly.  
There are two main conclusion to be drawn from this comparison:

(i) The $\eta\,^3$H interaction generated by 
the optical potential $U_I$ (\ref{eq165}) is relatively weak. 
It is interesting to compare our result with that obtained
within the scattering length approximation $t_{\eta N}\to
-\frac{2\pi}{\mu_{\eta N}}a_{\eta N}$. The latter predicts a  
binding of the $\eta\,3N$ system already for relatively modest values of 
$a_{\eta N}$ (see e.g.~\cite{Wilk93,HaHi99}). 
The trivial source of this discrepancy lies  
in the strong energy dependence of the $\eta N$ amplitude which is 
ignored by the scattering length approximation. 
The change of the free $\eta N$ energy $w_{\eta N}^{\mbox{\tiny free}}$ 
in the medium is primarily due to the Fermi motion and  
due to the binding of the nucleons. 
A rough estimation at zero $\eta\,^3$H kinetic energy gives 
\begin{equation}\label{eq180}
\Delta \omega=
w_{\eta N}-w_{\eta N}^{\mbox{\tiny free}}\approx -\varepsilon_b
-\frac{\langle q^2\rangle}{2M_3}\,,
\end{equation}
where $\varepsilon_b\approx 6.5$\,MeV is the binding 
energy of a participating nucleon to the two-nucleon core, while 
$\langle q^2\rangle$ stands for the mean squared 
nucleon momentum inside the nucleus,  
and the reduced mass $M_3$ is given by (\ref{19}).
Taking $\sqrt{\langle q^2\rangle}=120$ MeV/c we obtain  
$\Delta\omega\approx -15$ MeV. In the calculation, the energy 
at which the $\eta N$ amplitude has to be calculated was chosen 
according to the so-called "spectator on-shell" prescription.
The corresponding energy shift $|\Delta\omega|$ is larger than that given by 
the estimation (\ref{eq180}) where   
the internal energy of the two-nucleon core is neglected. Taking into account 
the difference $\Delta\omega$ results in decreasing the $\eta N$ scattering
amplitude and especially its imaginary part, which has a 
sharp energy dependence
around the $\eta N$ threshold (see Fig.~\ref{fig2}). 
The crucial importance of this fact was also discussed in \cite{Wyc95,Hai02}. 
One sees in Fig.~\ref{fig3} that  
inclusion of the intermediate nuclear interaction (the dash-dotted curve)
accounts for an appreciable 
portion of the noted 
disagreement, and it leads to a much better description near  
zero energy. We consider this fact as evidence that calculations,
which investigate the dependence of the $\eta A$-dynamics on the elementary
amplitude $t_{\eta N}$ but disregard the interaction of the 
participating nucleon with the surrounding nucleons, are of little 
significance. 

(ii) Comparison of the results obtained within the coherent approximation 
(\ref{eq135}) with the four-body treatment shows the role 
of higher order terms in the expansion (\ref{eq127}). 
As was already noted, their contribution is associated with 
virtual target excitations in between scatterings. As one sees, this effect 
is significant and increases as the energy approaches the 
inelastic threshold (analogous conclusions with respect to the $\eta
d$ interaction are given in \cite{Wyc01}). With increasing energy the 
cross section becomes similar to the one of the impulse approximation 
but as one sees in 
Fig.~\ref{fig3} the Argand plots remain very different. 

In our opinion, the conclusions above 
have an important bearing on models of the 
$\eta$-nuclear interaction. In particular, they point to the fact that such
models should not be developed as a mere repetition of the first-order 
$\pi$-nuclear scattering formalism. 

Returning to Fig.~\ref{fig3} we would like to note a strong enhancement 
of the cross section close to zero energy as a consequence of the 
$\eta\,^3$H virtual state. The scattering length 
$a_{\eta^3\!H}=(1.82+i2.75)$\,fm locates the position of the pole at 
$E_{\eta^3\!H}^{pole}\approx -1/(a^2_{\eta^3\!H}\mu_{\eta^3\!H})$
 = (1.53+i3.59) MeV. The pole lies on the first 
nonphysical sheet (${\cal I}m \sqrt{E}<0$) attached to the physical one 
through the two-body cut beginning at $\eta\,^3$H 
threshold~\cite{footn2}. The somewhat unusual behavior of the Argand plots 
near the inelastic threshold supposedly can be ascribed to 
a cusp-like structure of the amplitude with a rapidly varying real part.

In order to investigate the role of the short-range nucleon-nucleon dynamics
we have performed in addition a four-body calculation with a Yamaguchi
parametrization of the potential $v_{NN}$ \cite{Yam54} where the 
complicated structure of the short-range $NN$ interaction is ignored. 
The respective result is represented by 
the long-dashed curve in the Fig.~\ref{fig3}.
As one notes, the difference is insignificant. 
An obvious conclusion, which follows, is that in the low-energy region 
only the long range part of the $NN$ interaction comes into play, 
which may be described quite satisfactorily by the Yamaguchi
potential. In other words, our results are not sensitive to the 
$NN$ interaction 
models (which must, of course, be on-shell equivalent at low energy)
as long as the momenta in question are essentially smaller than those
associated with the short-range part of the $NN$ force.


\section{$\eta$-photoproduction on A=3 nuclei near threshold}\label{sect3}

\begin{figure}
\includegraphics[scale=.7]{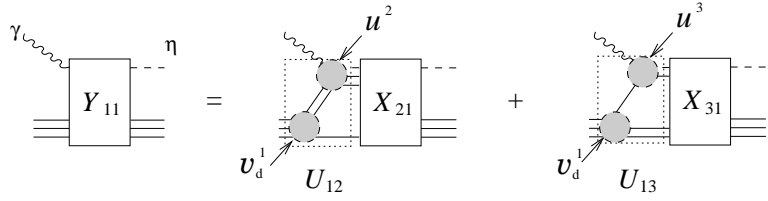}
\caption{
The photon induced effective potentials appearing in leading order 
of the electromagnetic interaction in $\eta$-photoproduction on three-body  
nuclei (\protect\ref{eq195}).
}
\label{fig4}
\end{figure}

Turning now to $\eta$-photoproduction, we treat the electromagnetic
interaction as usual up to the first order in the fine structure constant. 
As a consequence, the photon appears only in the initial state 
as an incident particle. This scheme is illustrated in Fig.~\ref{fig4} 
where the electromagnetic 
vertex functions $u^\alpha$ ($\alpha=2,3$) are of first
order in the $\gamma N$ coupling. 
The corresponding expression of the amplitude $Y^{(ss')}_{11}$ reads  
\begin{equation}\label{eq190}
Y^{(ss')}_{11}(p,k_\gamma;E)=\sum_{\alpha=2,3}\sum
\limits_{l,m}\sum_{\sigma=0,1}
\int\limits_0^\infty
X^{(s\sigma)}_{\alpha 1;l1}(p,p';E)
\,\Theta^{(\sigma)}_{\alpha;lm}\Big(E-\frac{{p'}^2}{2M_\alpha}\Big)
U^{(\sigma s')}_{1\alpha;1m}(p',k_\gamma;E)
\,\frac{{p'}^2dp'}{2\pi^2}\,,
\end{equation}
where the hadronic amplitudes $X_{\alpha 1}$ are defined in 
section~\ref{sect1} (see Eq.\,(\ref{eq10})).
Here the spin coupling in the initial $\gamma 3N$ state is
also uniquely determined by the spin $s$ of the $NN$-pair  
since the spin of the target is fixed to $S=1/2$. 
The effective potentials, involving the photon-induced excitation of the
resonance $N^*$, are defined by 
\begin{equation}\label{eq195}
U^{(ss')}_{1\alpha;nn'}(p,k_\gamma;E)=
\frac{\big(\Omega^{(t_\gamma)}_\alpha\big)_{ss'}}{2}
\int\limits_{-1}^{+1}v^{1(s)}_{d;n}(q_1,-\varepsilon_{^3\!H})
\,\tau^{(s)}_d\big(-\varepsilon_{^3\!H}-\frac{3q^2_1}{4M_N}\big)
u_{n'}^{\alpha(ss',t_\gamma)}(q_\alpha)
d(\hat{k}_\gamma\cdot\hat{p})\,, \quad \alpha=2,3\,,
\end{equation}
with a relative momentum at the $(3N)\to N+(NN)$ vertex 
$\vec{q}_1=\vec{p}+\frac{1}{3}\vec{k}_\gamma$. 
The form factors $u_n^{\alpha(ss',t_\gamma)}(q_\alpha)$ 
are associated with the absorption of a photon having isospin $t_\gamma$
by a nucleon or by a nucleon pair (see Fig.~\ref{fig4}).
For the $\gamma (NN)$ and $\gamma N$ relative momenta 
$q_\alpha$ $(\alpha=2,3)$ we use semirelativistic expressions  
\begin{equation}\label{eq198}
\displaystyle
\vec{q}_2=\vec{k}_\gamma +\frac{\omega_\gamma}{\omega_\gamma+2M_N}\ \vec{p}\,, 
\quad 
\vec{q}_3=\vec{k}_\gamma +\frac{\omega_\gamma}{\omega_\gamma+M_N}\ \vec{p}\,. 
\end{equation}

The spin-isospin coefficients, presented in (\ref{eq195}) in matrix
form by $\Omega^{(t_\gamma)}_\alpha$, are obtained from standard 
spin algebra
\begin{equation}\label{eq230}
\Omega_2^{(0)}=\left(
\begin{array}{cc}
0 & 0 \\
 & \\
0 & 2\sqrt{\frac{2}{3}} \end{array}
\right)\,, \quad
\Omega_2^{(1)}=\left(
\begin{array}{cc}
0 & \pm\sqrt{\frac{2}{3}} \\
 & \\
\pm\sqrt{\frac{2}{3}} & 0 \end{array}
\right)\,, \quad
\Omega_3^{(0)}=\left(
\begin{array}{cc}
\sqrt{6} & 0 \\
 & \\
0 & -\sqrt{\frac{2}{3}} \end{array}
\right)\,, \quad
\Omega_3^{(1)}=\left(
\begin{array}{cc}
\mp\sqrt{\frac{2}{3}} & 0 \\
 & \\
0 & \mp\sqrt{\frac{2}{3}} \end{array}
\right)
\end{equation}
with the upper (lower) sign referring to $^3$He ($^3$H), respectively.

It should be noted, that going from $\eta\,^3$H 
elastic scattering to $\eta$-photoproduction 
we are faced with qualitatively new physics
where large momentum transfers dominate. In particular, 
due to this reason, 
the contribution of pion exchange between the nucleons to the $\eta$
production mechanism must be included in general. 
This fact is confirmed by several theoretical
developments \cite{Liu92,RiAr00}. However, for reasons of principal 
numerical difficulties, already noted in Sect.~\ref{sect1},  
we do not include pion rescattering into our 
calculation and make only several remarks in the conclusion. 

As for the electromagnetic vertex functions $u^{\alpha}_{n}$, 
it is easy to show that up to the first order in the $\gamma
N$ interaction they are given by (cf.\ Eq.~(\ref{eq31}))
\begin{equation}\label{eq200}
u^{\alpha(ss',t_\gamma)}_{n}(q_\alpha)=\int\limits_0^\infty
\widetilde V^{(s,t_\gamma)}_{\alpha;dN^*}(q_\alpha,q')\,\tau_{N^*}
\big(B_\alpha-\frac{{q'}^2}{2\mu^\alpha_{N^*}}\big)
v^{\alpha(s')}_{N^*;n}(q',B_\alpha)
\frac{{q'}^2dq'}{2\pi^2}\,, \quad \alpha=2,3\,,
\end{equation}
where the effective potentials 
$\widetilde V^{(s,t_\gamma)}_{\alpha;dN^*}$ are determined 
analogously to the hadronic potentials $V_{\alpha;dN^*}$~\cite{FiAr02} but 
with an incident $\eta$ replaced by a photon
\begin{eqnarray}\label{eq205}
\widetilde 
V^{(s,t_\gamma)}_{2;dN^*}(q_2,q')&=&-\frac{1}{\sqrt{2}}\int\limits_{-1}^{+1}
\frac{
g_d^{(s)}\left(\left|\vec{q}\,'+\frac{1}{2}\vec{q}_2\,\right|\right)
\tilde g^{(t_\gamma)}_{N^*}(|\vec{q}_2 
+\frac{\omega_\gamma}{\omega_\gamma+M_N}\vec{q}\,'|,\omega_{N^*})}
{
\varepsilon_{^3\!H}+\frac{3}{4M_N}(\vec{p}+\frac{1}{3}\vec{k}_\gamma)^2
+\frac{1}{M_N}\left(\vec{q}\,'+\frac{1}{2}\vec{q}_2\,\right)^2}
\ d(\hat{q}\cdot\hat{q}')\,, \\
&&\nonumber \\
\widetilde V^{(s,t_\gamma)}_{3;dN^*}(q_3,q')\label{eq205a}
&=&
-\frac{g_d^{(s)}(q')\,
\tilde g^{(t_\gamma)}_{N^*}(q_3,\omega_{N^*})}
{
\varepsilon_{^3\!H}+\frac{3}{4M_N}(\vec{p}+\frac{1}{3}\vec{k}_\gamma)^2
+\frac{1}{M_N}{q'}^2}\,,
\end{eqnarray}
where $\tilde g^{(t_\gamma)}_{N^*}(k_{\gamma N},\omega_{N^*})$ denotes 
the $\gamma N\to N^*$ vertex functions depending on the $\gamma N$ c.m.\ 
momentum $k_{\gamma N}$ and the invariant $\eta N$ energy $\omega_{N^*}$. 
The denominators in the expressions (\ref{eq205}) and (\ref{eq205a})
are obtained by taking into account the on-shell conditions in the initial 
$\gamma\,^3$H state, i.e.\ $E=-\varepsilon_{^3\!H}+\omega_\gamma-m_\eta
+k^2_\gamma/2M_{^3\!H}$ as well as the dependence of the momenta $\vec{q}_2$
and $\vec{q}_3$ on $\vec{k}_\gamma$ and $\vec{p}$ given by (\ref{eq198}). 
One readily sees that the singularities on the real $q'$-axis are never reached
in $\widetilde V_{\alpha;ij}$. 

\begin{figure}
\includegraphics[scale=.7]{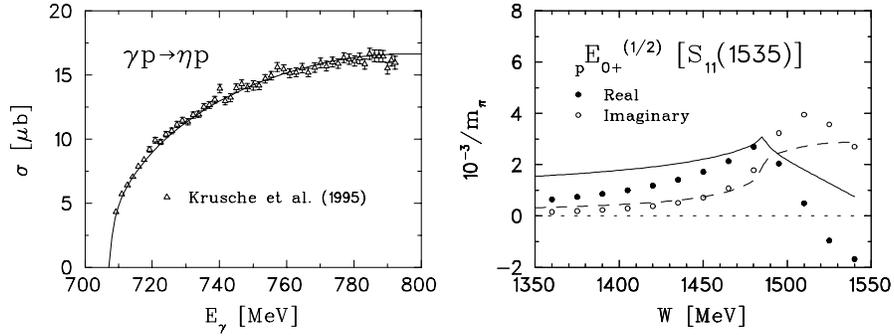}
\caption{Left panel: the $\gamma p\to\eta p$ total 
cross section compared 
with the data of~\protect\cite{Kru95}. 
Right panel:  
the $_pE_{0^+}^{(1/2)}$ multipole of $\gamma p\to\pi p$ generated by the 
photoexcitation of the $S_{11}(1535)$ resonance. Solid curve: real part, 
dashed curve: imaginary part. Open and filled circles 
represent the results of the MAID-parametrization~\protect\cite{MAID}. 
}
\label{fig5}
\end{figure}

In the actual calculation, we treat the vertices 
$\tilde g_{N^*}^{(t_\gamma)}(k,\omega_{N^*})$ independent of the momentum 
$k$ and parametrize their behavior in the following form 
\begin{eqnarray}\label{eq220}
\tilde g_{N^*}^{(1)}(k,\omega_{N^*})&=&\left\{
\begin{array}{ll}\displaystyle 
\frac{e}{\sqrt{4\pi}}
\sum\limits_{n=0}^4 a_n\left(\frac{k_\pi}{m_\pi}\right)^n, &
\quad \omega_{N^*}>M_N+m_\pi\,,\\
&\\
\tilde g_{N^*}^{(1)}(k,\omega_{N^*})
\Big|_{\omega_{N^*}=M_N+m_\pi}\,, & \quad \mbox{else}\,,
\end{array}
\right.\\
&&\nonumber\\
\tilde g_{N^*}^{(0)}(k,\omega_{N^*})&=&
0.1\,\tilde g_{N^*}^{(1)}(k,\omega_{N^*})\,,
\nonumber 
\end{eqnarray}
where $k_\pi$ is the on-shell pion momentum in the $\pi N$ c.m.\ frame 
corresponding to the total energy $\omega_{N^*}$. 
The isospin separation of the $S_{11}(1535)$
photoexcitation amplitude is chosen such that the relation
\begin{equation}\label{eq222}
\frac{\sigma(\gamma p\to\eta p)}
{\sigma(\gamma n\to\eta n)}= 0.67\,,
\end{equation}
is reproduced in accordance with the experimental results for 
quasifree $\eta$-photoproduction on light nuclei~\cite{Hoff97,Hej99}.
The coefficients in (\ref{eq220}) 
\begin{equation}
a_0=5.007\cdot 10^{-1}\,,\ 
a_1=-1.750\cdot 10^{-2}\,,\ 
a_2=0.926\cdot 10^{-1}\,,\ 
a_3=2.052\cdot 10^{-3}\,,\ 
a_4=-6.408\cdot 10^{-3}
\end{equation}
were obtained by fitting the $\gamma p\to \eta p$ data 
\cite{Kru95} as shown in Fig.~\ref{fig5}. 
In the same figure we also compare our calculation of the 
$\gamma p\to S_{11}(1535)\to \pi p$ amplitude $_pE^{(1/2)}_{0+}$ with the 
results of the MAID parametrization \cite{MAID}.
For definiteness, we present here the elementary photoproduction amplitude 
for $\gamma p\to \pi p$ 
\begin{equation}\label{eq223}
t_{\lambda}
=\ _pE^{(1/2)}_{0+}\,\left(\vec{\sigma}\cdot\vec{\varepsilon}_\lambda\right)
\ \quad 
\mbox{with}\ \ 
_pE^{(1/2)}_{0+}=
\big[\tilde g_{N^*}^{(0)}(k_\gamma,\omega_{N^*})+
\tilde g_{N^*}^{(1)}(k_\gamma,\omega_{N^*})\big]
\tau_{N^*}\big(\,\omega_{N^*}-M_N-m_\eta\big)
g_{N^*}^{(\eta)}(k_\eta)\,.
\end{equation} 
The c.m.\ differential cross section then reads 
\begin{equation}\label{eq224}
\frac{d\sigma}{d\Omega}(\gamma p\to\eta p)=\frac{k_\eta}{k_\gamma}\,
\frac{2\omega_\eta E_{N_i}E_{N_f}}{(4\pi\omega_{N^*})^2}\,
\left|\, _pE^{(1/2)}_{0+}\right|^2\,,
\end{equation} 
with $\omega_\eta$ and $E_{N_{i(f)}}$ denoting the energies of the 
$\eta$ meson and the initial (final) nucleon, respectively.

\begin{figure}
\includegraphics[scale=.7]{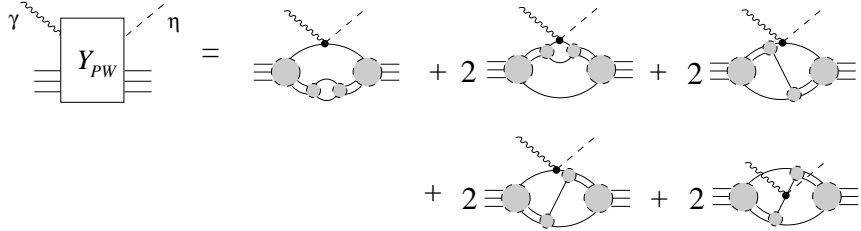}
\caption{
Schematic representation of the PWIA-term in the $\eta$-photoproduction 
amplitude (\protect\ref{eq235}) related to our model of the target wave 
function (\protect\ref{eq70})-(\protect\ref{eq74}). The factors 2 stem 
from the identity of the nucleons. 
}
\label{fig6}
\end{figure}

Before completing the formal part, we recall once more that all the
expressions above relate only to  
the $s$-wave. With increasing energy, higher
partial waves, where however no significant 
interaction is expected, are needed 
to fill the available phase space. To take into account their
contribution we use here the standard prescription 
\begin{equation}\label{eq235}
Y=Y_{PW}+\left[\,Y-Y_{PW}\right]_{L=0}\,,
\end{equation}
where $Y_{PW}$ is the plane wave approximation to the production 
amplitude. Assuming that the hadronic interaction in the higher partial
waves is insignificant, the difference in the parenthesis is reduced to
$s$-waves only. The amplitude $Y_{L=0}$ is given by (\ref{eq190}).
The diagrammatic representation of the amplitude $Y_{PW}$ is presented in the
Fig.~\ref{fig6}. 
The corresponding analytic expression is 
easily obtained and need not be presented here. We note only
that each term in the sum is represented by a 6-dimensional integral,
which were calculated numerically without any approximation. 

The reaction matrix element for the transition between the nuclear states 
with spin 1/2 is related to the amplitude (\ref{eq235}) 
by the Wigner-Eckart formula 
\begin{equation}\label{eq237}
\langle \frac{1}{2}M_f|\,T_\lambda|\frac{1}{2}M_i\rangle
=\frac{1}{\sqrt{2}}\,
(\frac{1}{2}M_i\,1\lambda\,|\frac{1}{2}M_f)\sum\limits_{ss'=0,1}
Y^{(ss')}\,.
\end{equation}
For the unpolarized c.m.\ cross section, we obtain
\begin{equation}\label{eq238}
\frac{d\sigma}{d\Omega}(\gamma A\to\eta A)=\frac{k_\eta}{k_\gamma}\,
\frac{2\omega_\eta E_{A_i}E_{A_f}}{(4\pi W)^2}\,
\frac{1}{6}
\big|\!\sum\limits_{ss'=0,1}
Y^{(ss')}(\vec{k}_\gamma,\vec{k}_\eta\,)\,\big|^2\,,
\end{equation}
with $E_{A_{i(f)}}$ 
being the total target energy in the initial (final) state. 

\begin{figure}
\includegraphics[scale=.7]{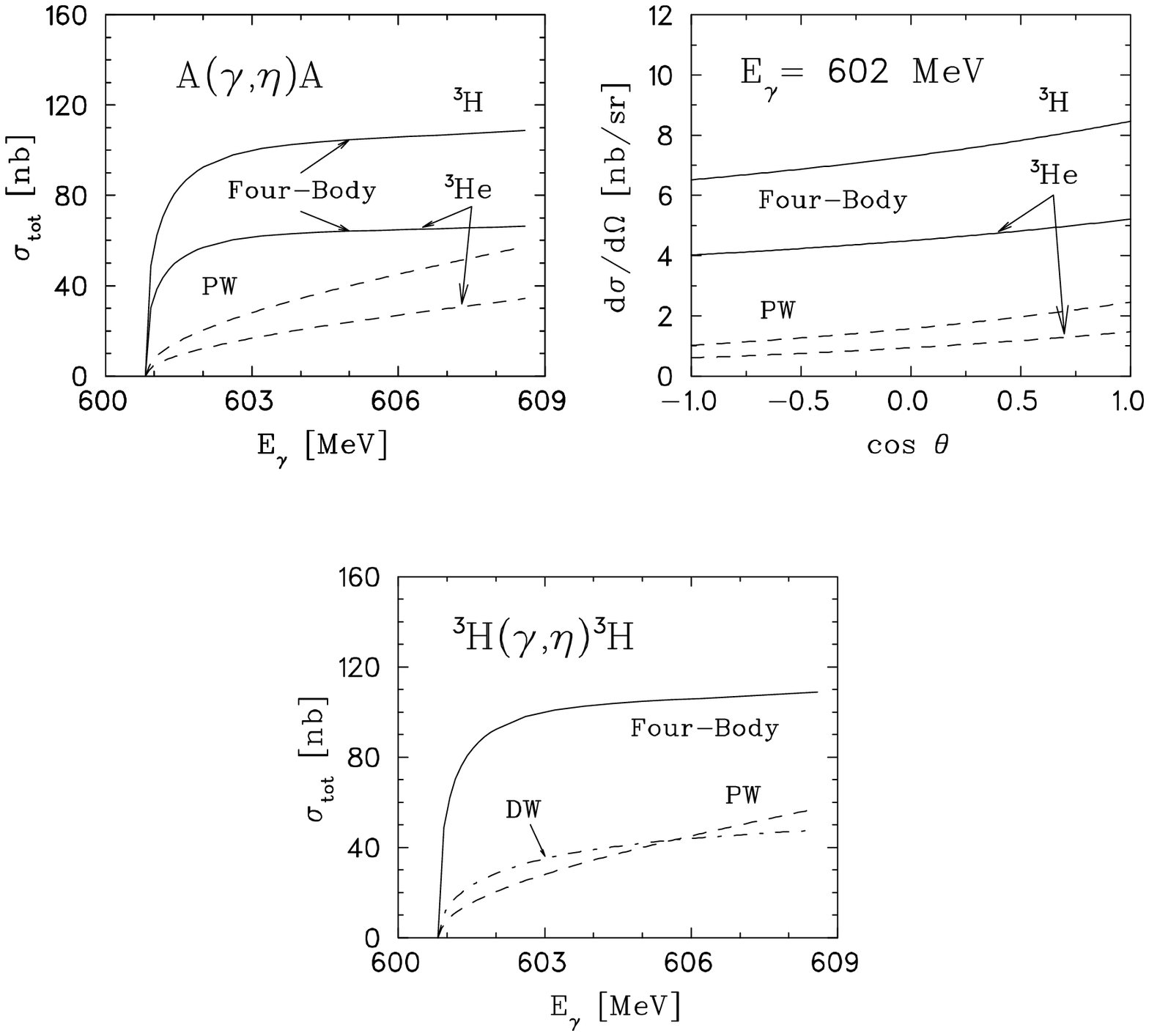}
\caption{Upper panels:  
total and differential cross section for $\eta$-photoproduction 
on $^3$He and $^3$H calculated within the four-body scattering model 
for the final $\eta\,3N$ system compared to the results of the 
plane-wave calculation (the dashed curves on both panels). 
In the lower panel the FSI effects provided by the distorted wave (DW) 
approach with the optical potential (\protect\ref{eq165}) (dash-dotted curve)
are compared with those given by the four-body calculation and the plane-wave 
approximation (PW). 
}
\label{fig7}
\end{figure}

Our predictions for total as well as differential cross sections  
are shown in Fig.~\ref{fig7}.   
Firstly we note an approximate equality  
\begin{equation}\label{eq240}
\frac{\sigma(\gamma n\to\eta n)}{\sigma(\gamma p\to\eta p)}\approx
\frac{\sigma(\gamma\, ^3\mbox{He}\to\eta\, ^3\mbox{He})}
{\sigma(\gamma\, ^3\mbox{H}\to\eta\, ^3\mbox{H})}
\approx 0.6\,.
\end{equation}
As was already explained in \cite{Shev02,TiBe94} this result is a 
consequence of the
spin-flip nature of the $\eta$-photoproduction amplitude $t_\lambda$ 
(\ref{eq223}). Approximating
the spatial part of the target wave function (\ref{eq70}) by only the 
principal,  
totally symmetric $s$-state it is easy to see that the 
$\eta$-meson can only be 
produced on the neutron in $^3$He and on the proton in $^3$H. 
The remaining two nucleons, coupled to a total spin $s = 0$, 
do not participate due to the Pauli principle. 
The $\eta$-rescattering effects, being spin-independent do not 
distort this relation. A small deviation from the 
relation (\ref{eq222}) is simply due to the presence of the 
state with mixed permutation symmetry. 

As expected, the final state interaction leads to a 
rather pronounced enhancement of the plane wave result, especially very close 
to the production threshold. The cross section 
reaches very fast its characteristic value and has a form of a flat plateau. 
The angular distribution of $\eta$-mesons in both reactions 
is shown in the upper 
right panel of the Fig.~\ref{fig7}. Within our model only the
angular-independent $s$-wave part of the $\eta\,3N$ wave function undergoes
distortion due to the FSI. As a consequence,  
the differential cross section is much more isotropic as 
compared with the plane wave calculation.

Comparing our results for $^3$He$(\gamma,\eta)^3$He reaction 
with those of Ref.~\cite{Shev02} obtained within the
finite-rank-approximation (FRA), 
we observe rather well agreement in magnitude of the total cross sections
close to the threshold 
(we take for comparison the results of \cite{Shev02} corresponding to 
the model IIIa for the $\eta N$-interaction).
However we think, this fact has no physical significance, since   
there are principal differences in the models.  
Firstly, we would like to note a disagreement concerning the nature of the 
final state interaction. Namely, as explained in Ref.~\cite{Shev02} the 
strong effect of FSI, found by the authors, is due to the $s$-wave 
$\eta\,3N$ resonance, located near zero kinetic energy \cite{Belya95}. 
In our case it is a consequence of the virtual state, as was already 
discussed in \cite{FiAr02}. We do not find any evidence for the resonance 
behavior of the $\eta\,^3$H amplitude 
(see, e.g., the Argand plots in Fig.~\ref{fig3}).
Furthermore, our calculation does not exhibit a strong slope in the 
cross section caused by a cusp at the inelastic threshold, 
which was found to be very pronounced in~\cite{Shev02}. 

In the lower panel of Fig.~\ref{fig7}, we also depict the total cross section 
for the reaction on $^3$H where the final state is distorted by the 
first-order optical potential (the approximation denoted as DW).   
The corresponding photoproduction amplitude is given by  
(cf.\ Eq.~(\ref{eq175}))
\begin{equation}
Y(\vec{k}_\eta,\vec{k}_\gamma;E)=Y_{PW}(\vec{k}_\eta,\vec{k}_\gamma;E)
+\frac{2}{3}\,\frac{\mu_{\eta\,^3\!H}}{\pi^2}\int\limits_0^\infty
\frac{T(k_\eta,p;E)\ [Y_{PW}(\vec{p},\vec{k}_\gamma;E)]_{L=0}}
{k_\eta^2-p^2+i\varepsilon}\ {p}^2 dp\,,
\end{equation}
where the $T$-matrix for the $\eta\,^3$H scattering $T(k_\eta,p;E)$ was 
calculated with the potential (\ref{eq165}). 
As one readily notes, the DW approach visibly underestimates the strong 
FSI effect of the four-body theory. 

A comparison of the DW calculation for the reaction 
$^3$He$(\gamma,\eta)^3$He with 
the full model has also been done in~\cite{Shev02}. In 
particular, the authors have noted a very large difference between the cross 
sections obtained within FRA and DW approaches. At the energy 
$E_\gamma$ = 605\,MeV the DW results reported in \cite{Shev02} 
underpredict those of the FRA model by about a factor 20. 
The reason of this disagreement
is a very strong suppression of the DW cross section in the near-threshold 
region. In contrast to this conclusion, 
our calculation predicts a typical $s$-wave energy 
dependence of the cross section for the coherent reaction of the form  
\begin{equation} 
\sigma\sim\sqrt{E_\gamma - E_\gamma^{th}}\,,
\end{equation} 
with $E_\gamma^{th}$ denoting the threshold energy. This form is slightly 
distorted by the $\eta$-nuclear optical potential which tends 
to increase the cross section value close to the threshold.
As a consequence the 
difference between the DW result and the full four-body calculation 
turns out to be not so impressive as in \cite{Shev02}.

\section{Conclusion}

In the present paper we have investigated elastic
scattering and photoproduction of $\eta$-mesons on three-body nuclei
near threshold. The possibility of having the exact solution at hand 
permits us to investigate unambiguously the corrections to the 
lowest-order optical potential which are usually neglected within
the "standard" optical model approach. 
According to the results presented above, we would like to 
draw the following conclusions:

(i) The $\eta N$ scattering amplitude 
is appreciably modified in the nuclear medium. 
The contributions beyond the impulse approximation
turn out to be very important. It is reasonable to assume that 
the origin of this fact lies in the 
resonance nature of the $\eta N$ amplitude 
giving rise to large corrections caused by the binding of the nucleons. 
One may expect that this effect is even stronger in heavier nuclei.

(ii) The influence of virtual target excitations between successive 
scatterings is also rather important. 
Although the three- and four-body thresholds 
are relatively far from the $3N$ binding energy, neglect of the excited 
states makes the result very different from the exact one. 
In other words, the contributions of virtual three- and four-body states 
are also quite important below the corresponding unitary cuts. 

(iii) Since in the energy region considered here the incident energy 
of the $\eta$-meson remains small, i.e., its
wave-length is large compared to the characteristic internuclear distance, 
the results for the $\eta\,^3$H scattering are not visibly sensitive to 
the details of the short-range $NN$-dynamics. 
This conclusion is confirmed straightforwardly comparing the results of the 
PEST potential~\cite{Zan83}  
with those given by the simplest Yamaguchi form of the $NN$ interaction. 

(iv) Close to the threshold, the final state 
interaction enhances the $\eta$-yields appreciably, which was already
noted in a variety of studies of $\eta$-production on lightest
nuclei with different entry channels~\cite{May96,Cal97,Hej02}. 
The angular distribution shows pronounced 
isotropy, associated with the $s$-wave dominance of FSI. 

In conclusion, we would like to note once more the possible importance
of pion exchange in the $\eta$-photoproduction on nuclei. 
One can expect this since 
the suppression due to the strong momentum transfer which is
presumably important for low-energy $\eta\,3N$ scattering appears not to be 
effective here. Furthermore, as was already noted, due to the spin-flip nature
of the $\eta$-photoproduction mechanism, only $\approx$\,1/3 
of the nucleons are involved in the process. 
This may further enhance the importance of the $\pi$-exchange
contribution where the nonvanishing non-spin-flip part gives rise to a 
coherent enhancement of the reaction strength. A good case in point is 
the pion production via $\Delta(1232)$ excitation (the
spin-independent part dominates) with subsequent rescattering into $\eta$
through the excitation 
of $S_{11}(1535)$ (the spin-flip part is negligible) on the next nucleon.
On the other hand, we suppose that due to the short range nature of the
pion-rescattering 
mechanism its contribution does not influence the strong energy dependence of
the cross section discussed above but its magnitude can be visibly  
affected.

\acknowledgments
The work was supported by the Deutsche Forschungsgemeinschaft (SFB 443).



\begin{thebibliography}{99}

\bibitem{Ued91}
T. Ueda, Phys. Rev. Lett. {\bf 66}, 297 (1991), 
Phys. Lett. B {\bf 291}, 228 (1992).

\bibitem{Wilk93}
C. Wilkin, Phys. Rev. C {\bf 47}, R938 (1993).

\bibitem{Wyc95}
S. Wycech, A.M. Green, J.A. Niskanen, Phys.\ Rev.\ C {\bf 52}, 
544 (1995).

\bibitem{FiAr02}
A. Fix and H. Arenh\"ovel, Phys.\ Rev.\ C {\bf 66}, 024002 (2002).

\bibitem{ErEr66}
M. Ericson and T.E.O. Ericson, Ann.\ Phys.\ {\bf 36}, 323 (1966).

\bibitem{Pfeif02}
M. Pfeiffer, Ph.D.\ Thesis (Giessen 2002).

\bibitem{Shev02}
N.V. Shevchenko {\it et al.}, Nucl.\ Phys.\ A {\bf 699}, 165 (2002).

\bibitem{Sof82}
S.A. Sofianos, N.J. McGurk, and H. Fiedeldey,  Nucl.\ Phys.\ A {\bf
318}, 295 (1979).

\bibitem{Fons86}
A. C. Fonseca, Proc.\ 8th Autumn 
School on Models and Methods in Few-Body Physics, Lisboa, 1986, eds. 
L.S. Ferreira, A.C. Fonseca
and L. Streit, Lecture Notes in Physics {\bf 273}, 161 (1986).

\bibitem{footn1}
There are separable expansion methods, which seem to be 
capable of approximating the three-body off-shell scattering amplitudes 
also at positive energies (see e.g.~A.C. Fonseca, H. Haberzettl, and 
E. Cravo, Preprint BONN-HE-82, July 1982). But they require far more 
numerical efforts and were not adopted here.

\bibitem{Zan83}
H. Zankel, W. Plessas, and J. Haidenbauer, 
Phys.\ Rev.\ C {\bf 28}, 538 (1983).

\bibitem{Hai02}
Q. Haider and L.C.\ Liu, Phys.\ Rev.\ C {\bf 66}, 045208 (2002).

\bibitem{BeTa90}
C. Bennhold and H. Tanabe, Nucl.\ Phys.\ A {\bf 530}, 625 (1991).

\bibitem{Arn85}
R.A. Arndt, J.M. Ford, and 
L.D. Roper, Phys.\ Rev.\ D {\bf 32}, 1085 (1985).

\bibitem{BSSN95}
M. Batini$\acute{\mbox{c}}$, I. $\check{\mbox{S}}$laus, 
A. $\check{\mbox{S}}$varc, and B.M.K. Nefkens, Phys.\ Rev.\ C 
{\bf 51}, 2310 (1995).

\bibitem{GoWa64}
M.L. Goldberger and K.M. Watson, 
{\it Collision Theory} (Wiley, N.Y., 1964).

\bibitem{KMT59}
A.K. Kerman, H. McManus, and R.M. Thaler, Ann.\ Phys.\ (N.Y.) 
{\bf 8}, 551 (1959).

\bibitem{AAY64}
R. Aaron, R.D. Amado, and Y.Y. Yam, Phys.\ Rev.\ {\bf 136}, B650 (1964).

\bibitem{Wyc01}
S. Wycech and A.M.\ Green, Phys.\ Rev.\ C {\bf 64},
045206 (2001).

\bibitem{NoKo65}
H.P. Noyes, Phys.\ Rev.\ Lett.\ {\bf 15}, 538 (1965);  
K.L. Kowalski, Phys.\ Rev.\ Lett.\ {\bf 15}, 798 (1965).

\bibitem{HaHi99}
R.S. Hayano, S. Hirenzaki, and A. Gillitzer, Eur.\ Phys.\ J.\ A
{\bf 6}, 99 (1999).

\bibitem{footn2}
More exactly, there are two poles which are placed symmetrically 
about the real energy axis.

\bibitem{Yam54}
Y.\ Yamaguchi, Phys.\ Rev.\ {\bf 95}, 1628 (1954).

\bibitem{Liu92}
L.C.\ Liu, Phys.\ Lett.\ B {\bf 288}, 288 (1992).

\bibitem{RiAr00}
F. Ritz and H. Arenh\"ovel, Phys.\ Rev.\ C {\bf 64}, 034005 (2001). 

\bibitem{TiBe94}
L. Tiator, C. Bennhold, and S.S. Kamalov, 
Nucl.\ Phys.\ A {\bf 580}, 455 (1994).

\bibitem{Hoff97}
P. Hoffmann-Rothe {\it et al.}, Phys.\ Rev.\ Lett.\ {\bf 78}, 4697 (1997).

\bibitem{Hej99}
V. Hejny {\it et al.}, Eur.\ Phys.\ J.\ A {\bf 6}, 83 (1999).

\bibitem{Kru95}
B. Krusche {\it et al.}, Phys.\ Rev.\ Lett.\ {\bf 74}, 3736 (1995).

\bibitem{MAID}
D. Drechsel, O. Hanstein, S.S. Kamalov, and L. Tiator, 
Nucl.\ Phys.\ A {\bf 645}, 145 (1999).

\bibitem{Belya95}
V.B.\ Belyaev, S.A.\ Rakityansky, 
S.A.\ Sofianos, M.\ Braun, and W.\ Sandhas,
Few Body Syst., Suppl.\ {\bf 8}, 309 (1995); 
S.A.\ Rakityansky, S.A.\ Sofianos, M.\ Braun, 
V.B.\ Belyaev, and W.\ Sandhas,
Phys. Rev.\ C {\bf 53}, R2043 (1996).

\bibitem{May96}
B. Mayer {\it et al.}, Phys.\ Rev.\ C {\bf 53}, 2068 (1996).

\bibitem{Cal97}
H. Calen {\it et al.}, Phys.\ Rev.\ Lett.\ {\bf 79}, 2642 (1997).

\bibitem{Hej02}
V. Hejny {\it et al.}, Eur.\ Phys.\ J.\ A {\bf 13}, 493 (2002).

\end{thebibliography}
\end{document}